\documentclass[12pt]{article}
\usepackage{amsfonts}
\usepackage{amssymb}
\usepackage{amsmath}

\newcommand\be{\begin{equation}}
\newcommand\ee{\end{equation}}
\newcommand\bea{\begin{eqnarray}}
\newcommand\eea{\end{eqnarray}}

\newcommand{\C}{{\mathbb C}}
\newcommand{\R}{{\mathbb R}}

\newcommand{\Td}{{{\mathbb T}^d}}
\newcommand{\Z}{{\mathbb Z}}

\newcommand{\lv}{l_\text{v}}
\newcommand{\Et}{E_\text{t}}
\newcommand{\lt}{l_\text{t}}
\newcommand{\lvxt}{l_{\text{v}\times\text{t}}}
\newcommand{\Evxt}{E_{\text{v}\times\text{t}}}
\newcommand{\EDensF}{E^{1/2}}
\newcommand{\lDensF}{l^{1/2}}

\newcommand{\EDensB}{E^{1}}
\newcommand{\lDensB}{l^{1}}

\newcommand{\Ed}{{\cal E}_d}
\newcommand{\ld}{l_d}
 \newcommand{\diag}{\text{diag}}
 \newcommand{\trace}{\text{Tr}}    
 \newcommand{\inv}{^{-1}}
 \newcommand{\tran}{^\text{t}}     
\newcommand{\cC}{{\cal C}}

\newcommand{\Cl}{\text{Cl}}
\newcommand{\Homeo}{\text{Homeo}}

\begin{document}
\begin{titlepage}
\centerline{\bf\Large  A New and Unifying Approach to Spin Dynamics}
\centerline{\bf\Large  and Beam Polarization in Storage Rings}
\vspace*{2em}
\centerline{\large K.~Heinemann\footnote{\tt heineman@math.unm.edu} \hspace*{1em} and \hspace*{1em}
                   J.A.~Ellison\footnote{\tt ellison@math.unm.edu}}
\centerline{\textit{Department of Mathematics and Statistics, University of New Mexico, Albuquerque, New Mexico, USA}}
\vspace*{2em}
\centerline{\large D.P.~Barber\footnote{{\tt mpybar@mail.desy.de}}$^,$\footnote{ 
                                         Also Visiting Staff Member at the Cockcroft Institute, Sci-Tech Daresbury,
                                              and at the University of Liverpool, UK}
                    \hspace*{1em} and \hspace*{1em}
                   M.~Vogt\footnote{\tt vogtm@mail.desy.de}}
\centerline{\textit{Deutsches Elektronensynchrotron DESY, Hamburg, Germany}}
\vspace*{2em}
\centerline{(Version 2, December 11, 2014)}

\begin{abstract}
With this paper we extend our studies \cite{BEH} on polarized beams by distilling tools from the theory of
 principal bundles. 
Four major theorems are presented,
 one which ties invariant fields with the notion of normal form,
 one which allows one to compare different invariant fields,
 and two that relate the existence of invariant fields to the existence of certain invariant sets
 and relations between them. 
We then apply the theory to the dynamics of
spin-$1/2$ and spin-$1$ particles and their
 density matrices describing statistically 
the particle-spin content of bunches.
Our approach thus unifies the spin-vector dynamics from
the T-BMT equation with the spin-tensor dynamics and other dynamics.
This unifying aspect of our approach relates the examples elegantly and
 uncovers relations between the various underlying dynamical systems in a 
transparent way.

\end{abstract}
\end{titlepage}

%
%
%
\section{\label{sec:intro}Introduction}
We continue our study \cite{BEH} of spin dynamics in storage rings and
extend our tool set by employing
a method developed in the 1980s by R. Zimmer, R. Feres 
and others for Dynamical-Systems theory \cite{Zi,Fe,HK}.
In contrast to \cite{BEH}, we now employ a discrete-time treatment
(continuous-time would do as well).

Polarization of spin-$1/2$ and spin-$1$ particles 
is best systematized in a statistical description in terms
of invariant spin fields (ISF's) and 
invariant polarization-tensor fields (IPTF's). They are essential for
estimating the maximum attainable proton and deuteron polarization. 
Moreover, the invariant spin field appears in estimates of the electron 
and positron
equilibrium polarization due to synchrotron radiation.  
While the conditions for the existence of invariant spin fields
are not well understood, 
there are at least four practical algorithms which compute
ISF's when they exist.
For example, there is a
perturbative method \cite{Mane} from the mid 1980's 
(implemented in the
code SMILE). Then there is a Fourier-method \cite{ky99} 
from the 1990's
(implemented in the codes SODOM and SODOM2). The ISF can also be 
computed via  
a normal-form algorithm based on
Truncated Power Series Algebra (TPSA) from the 1990's (implemented 
in the code COSY-Infinity \cite{B}
and in the PTC/FPP code \cite{F}). The perhaps most flexible   
algorithm  uses so-called stroboscopic averaging \cite{HH96}. This was invented in the mid-1990's
and was implemented in the code SPRINT \cite{mv2000}.
For mathematical details on stroboscopic averaging, see \cite{EH}.
Stroboscopic averaging can also be used to 
obtain the IPTF   \cite{BV2}.
For more details on these four algorithms, see \cite{BHV}.
The numerical evidence produced by the above mentioned codes indicates
that invariant fields can be rather complex entities.
Moreover the question of existence, although trivial
in some simple cases, is up to this day, unsolved and, as evidenced by
SPRINT and SODOM2, situations can occur where invariant spin fields 
might not exist. In other
words there is no good understanding of the conditions under
which the invariant spin field exists.
Moreover it would be useful to have alternative and more efficient
algorithms for computing invariant fields.
It is in the light of these facts that we now present a new approach, started in \cite{He}, to the understanding 
of invariant fields.

We present four major theorems,
 the Normal Form Theorem 
tying invariant fields with the notion of normal form,
 the Decomposition Theorem, allowing comparison of different invariant fields,
 the Invariant Reduction Theorem, giving new insights into 
the question of the existence of invariant fields
and which is supplemented by the Cross Section Theorem.
It turns out that the well-established notions of
invariant frame field, spin tune, and spin-orbit
resonance are generalized by the normal form concept whereas
the well-established notions of invariant spin field and
invariant polarization-tensor field are generalized to invariant
$(E,l)$-fields. 
Here the notation $(E,l)$ indicates an
$SO(3)$-space which is shorthand 
for $l$ being a continuous $SO(3)$-action on a topological space $E$, i.e.,
$l:SO(3)\times E\rightarrow E$ is continuous and
$l(I_{3\times 3};x)=x \; \mbox{and}\; l(r_1r_2;x)=l(r_1;l(r_2;x))$
where $I_{3\times 3}$ is the unit $3\times 3$ matrix. 
With the flexibility in the choice of $(E,l)$,
we have a unified way to study the dynamics of spin-$1/2$ 
and spin-$1$ particles and their bunch density matrices.
Accordingly several $(E,l)$'s are discussed in some detail.
The origins of our formalism, in bundle theory, are pointed out 
and we also mention the relation to Yang-Mills Theory.
We thus open a significant new area of research in our field by bringing in
techniques from Bundle Theory used hitherto in very different research
areas.
We believe that all four of
our theorems are important and new.
%
%
%
%
%
%
%
%
%
%
%
\section{\label{sec:GenPartSpinMotion} General Particle-Spin 
Dynamics}
We assume integrable particle motion so that for fixed
action the particle motion is on the $d$-torus $\Td$.
The particle one-turn map is given by
\be
 z\in\Td \mapsto j(z)\,,\, j\in \Homeo(\Td) \,.\label{eq:ParticleMap}
\ee
Here $\Homeo(\Td)$ denotes the set of homeomorphisms on $\Td$.
Now we add to each particle an additional $E$-valued ``spin-like'' quantity $x$.
The one-turn particle-``spin'' map is the function
${\cal P}[E,l,j,A]:\Td \times E \to \Td \times E$ defined by
\be
{\cal P}[E,l,j,A](z,x)=(j(z),\,l(A(z);x)) \; ,\label{eq:ParticleSpinMap}
\ee
 where $A\in \cC(\Td,SO(3))$ is the one-turn spin transfer matrix.
Here $\cC(X,Y)$ denotes the set of continuous functions
 from $X$ to $Y$ and $SO(3)$ is the group of real orthogonal $3\times 3$-matrices
 of determinant $1$.
In our formalism, (\ref{eq:ParticleSpinMap}) is the most
general description of particle-spin dynamics
and the choice of $(E,l)$ depends on the situation as we illustrate
in the examples.
\section{\label{sec:GenFieldMotion} General Field Dynamics}
We are primarily interested in the field dynamics induced by the particle-spin 
dynamics.
Let $f:\Td\rightarrow E$ be an $E$-valued field on $\Td$ and set 
$x=f(z)$ in (\ref{eq:ParticleSpinMap}).
Then, after one turn, $z$ becomes $j(z)$
and the field value at $j(z)$ becomes $l(A(z);f(z))$. 
Thus after one turn 
the field $f$ becomes the field 
$f':\Td\rightarrow E$ where
$f'(z):=l(A(j\inv(z));f(j\inv(z)))$.
Therefore we have the one-turn field map
\begin{eqnarray}
&& 
f\mapsto f'= l(A\circ j\inv;f\circ j\inv) \; ,
\label{eq:FieldMap}
\end{eqnarray}
where $\circ$ denotes composition.
Iteration of (\ref{eq:FieldMap}) gives the field after 
$n$ turns resulting
in a formula involving the $n$-turn spin transfer matrix.
We call $f\in \cC(\Td,E)$
an ``invariant $(E,l)$-field of $(j,A)$'', or just ``invariant field'',
if it is mapped by 
(\ref{eq:FieldMap}) into itself, i.e., 
\begin{eqnarray}
&&
f\circ j = l(A;f) \; .
\label{eq:StationaryEquation}
\end{eqnarray}
%
Our main focus is on exploring invariant fields
as these describe 
the spin equilibrium of a bunch \cite{BEH}.

We work in the framework of topological dynamics. So 
$A,j,l,f$ are
continuous functions. Frameworks using weaker or stronger conditions than
continuity are possible.
%
\section{\label{sec:examples} Examples of $SO(3)$-spaces $(E,l)$}
\subsection{Example 1: $(E,l)=(\R^3,\lv)$}
%
For spin-$1/2$ particles the spin variable is the spin-vector 
$x=S\in\R^3$. The T-BMT equation \cite{BEH} leads us to
describe the particle-spin
and field dynamics in terms of the
$SO(3)$-space $(\R^3,\lv)$ where 
$\lv(r;S):=rS$ and $r\in SO(3)$. 
Thus, by (\ref{eq:ParticleSpinMap}), 
${\cal P}[\R^3,\lv,j,A](z,S)=(j(z),\,A(z)S)$.
%
%
\subsection{Example 2: $(E,l)=(\Et,\lt)$}
The $SO(3)$-space $(\R^3,\lv)$ does not suffice for spin-$1$ particles.
One also needs
the spin variable $x=M\in\Et$, called the spin-tensor, where
$\Et:=\{M\in\R^{3\times3} : M\tran=M, \trace(M)=0\}$. 
The particle-spin tensor dynamics and corresponding
field dynamics is described by $(\Et,\lt)$ where
$\lt(r;M):=rMr\tran$ and $M\in E_t,r\in SO(3)$
and where the superscript $\tran$ denotes transpose.
Thus, by (\ref{eq:ParticleSpinMap}), 
${\cal P}[E_t,l_t,j,A](z,M)=(j(z),\,A(z)MA(z)\tran)$.
\subsection{\label{ssec:vDeMat} 
Example 3: $(E,l)=(\EDensF,\lDensF)$} 
%
The statistical description of a bunch of
spin-$1/2$ particles, in terms of density matrices, requires
the $SO(3)$-space $(\EDensF,\lDensF)$
where $\EDensF := \{ R \in {\mathbb C}^{2\times2} : R^\dagger = R, \trace(R)=1\}$, 
where $^\dagger$ denotes the hermitian conjugate
and $\lDensF$ is defined as follows. Let 
$\alpha\in\Homeo(\R^3,\EDensF)$ be defined by
\be
 \alpha(S) := \frac{1}{2}(I_{2\times2} +\sum_{i=1}^3 S_i\sigma_i) \;, 
\label{eq:VecToDensF}
\ee
where $S_i$ is the $i$-th component of $S$
and the $\sigma_i$ are the well-known Pauli matrices. Then
\bea
 \lDensF(r;\alpha(S)) & := & \alpha\left( \lv(r;S) \right) \,,
  \label{eq:SOThrActDensF}
\eea
i.e., $\alpha$ is a so-called ``$SO(3)$-map'' from 
$(\R^3,\lv)$ to $(\EDensF,\lDensF)$.
%
\subsection{\label{ssec:vXtDeMat}
Example 4: $(E,l)=(\EDensB,\lDensB)$}
%
To complete our treatment of spin-$1$ particles and their motions,
we must combine the vector and spin-tensor motion. 
The required $SO(3)$-space is $(\Evxt,\lvxt)$ where 
$\Evxt:=\R^3\times\Et$ and $\lvxt(r;S,M):=(\lv(r;S),l_t(r;M))=(rS,rMr\tran)$.

The statistical description of a bunch of spin-$1$ particles, in 
terms of density matrices, requires
the $SO(3)$-space $(\EDensB,\lDensB)$
where $\EDensB:=\{ R \in {\mathbb C}^{3\times3} : R^\dagger = R, \trace(R)=1\}$, 
and $l=\lDensB$ is defined as follows. Let 
$\alpha\in\Homeo(\Evxt,\EDensB)$ be defined by
%
\be
\alpha(S,M) = \frac{1}{3}\Big( I_{3\times 3} + \sum_{i=1}^3 S_i{\mathfrak J}_i 
              \sqrt{\frac{3}{2}}\sum_{i=1}^3 M_{ij}
                               ({\mathfrak J}_i{\mathfrak J}_j+{\mathfrak J}_j{\mathfrak J}_i)\Big) \hspace*{-0.5ex} \nonumber
\ee
where the hermitian $3\times 3$ matrices ${\mathfrak J}_i$ are given by
their nonzero components
${\mathfrak J}_1(1,2)={\mathfrak J}_1(2,3)=-i/\sqrt{2},
{\mathfrak J}_2(1,1)=-{\mathfrak J}_2(3,3)=1$ and
${\mathfrak J}_3(1,2)={\mathfrak J}_3(2,3)=1/\sqrt{2}$. Then
\bea
 \lDensB(r;\alpha(S,M)) & := & \alpha\left( \lvxt(r;S,M) \right) \,,
  \label{eq:SOThrActDensB}
\eea
i.e., $\alpha$ is an $SO(3)$-map from 
$(\Evxt,\lvxt)$ to $(\EDensB,\lDensB)$.
%
\section{\label{sec:FieldMotion}Invariant fields}
We now continue our general discussion.
Let $E_x:= l(SO(3);x) :=\lbrace l(r;x):r\in SO(3)\rbrace$. Then the $E_x$ partition $E$
and each set $\Td\times E_x$ is invariant under the particle-spin
motion of (\ref{eq:ParticleSpinMap}) 
and so we ``decompose'' $\Td\times E$.
Of central interest are invariant $(E,l)$-fields
in the situation when $E$ is Hausdorff and $j$ is
topologically transitive. The latter means that
a $z_0\in\Td$ exists such that the set
$\lbrace j^n(z_0):n=0,\pm 1,\pm 2,...\rbrace$ is dense in $\Td$, i.e., that its
closure equals $\Td$ and we then say for brevity that $j$ is $TT(z_0)$.
Remarkably, 
in this situation, every invariant $(E,l)$-field 
of $(j,A)$ takes values only in one $E_x$; the proof \cite{HBEV2}
uses the compactness of $SO(3)$.
Thus our main theorems below are formulated for fields
taking values in only one $E_x$ (though these theorems could be generalized).

In Example 1, a field
$f\in\cC(\Td,\R^3)$ is mapped, according
to (\ref{eq:FieldMap}), to $f'$ where
$f'(z)=A(j\inv(z))\,f(j\inv(z))$. Moreover
an invariant $(\R^3,\lv)$-field is also called an 
``invariant polarization field (IPF)''. Furthermore if
 $|f|=1$, it is called an ``invariant spin field (ISF)''.
Clearly $E_x=\lbrace S\in\R^3:|S|=|x|\rbrace$ is a sphere centered at $(0,0,0)$.
Thus if $j$ is topologically transitive and $f$ is an IPF of $(j,A)$
then $|f(z)|$ is independent of $z$.
The so-called ``ISF-conjecture'' states: If
$j$ is topologically transitive then an ISF exists.
\section{\label{sec:MainTheorems} Main Theorems}
%
\subsection{\label{ssec:NFT} The Normal Form Theorem (NFT)}
Let $T\in\cC(\Td,SO(3)),\; x\in E$ and $A'(z):=T\tran(j(z))A(z)T(z)$. 
Then
the NFT states
that $f(z):= l(T(z);x)$ is an invariant $(E,l)$-field of $(j,A)$
iff $l(A'(z);x))=x$ for all $z\in\Td$.
This leads to two important concepts in our work, the normal form and the isotropy group.
If $A'(z)$ above belongs to a subgroup $H$ of $SO(3)$ for all $z$ then we call $(j,A')$ an $H$-normal form of $(j,A)$. Moreover
the subgroup $H_x:=\{r\in SO(3) : l(r;x)=x\}$ is called the isotropy group
at $x$ (note that for every $H$ one can find $(E,l)$ and $x$ such that
$H=H_x$). Thus the NFT is reformulated as:
The function $f$ is an invariant 
$(E,l)$-field of $(j,A)$ iff $(j,A')$ is an $H_x-$normal form of $(j,A)$.
We emphasize that 
the NFT gives us a way to view invariant fields which
is distinctively different from
(\ref{eq:StationaryEquation}).

The normal form also impacts the notion of ``spin tune''.
By definition, $(j,A)$ has spin tunes iff
$T\in\cC(\Td,SO(3))$ exists such that $A'(z)$ is independent of $z$
\cite{HBEV2}.  
For $\nu\in[0,1)$, define  
$G_\nu:=\{R(2\pi n\nu):n\in\Z\}$
where the matrix $R(\mu)$ is given by its nonzero components
$R(\mu)(1,1)=R(\mu)(2,2)=\cos\mu,R(\mu)(2,1)=-R(\mu)(1,2)=\sin\mu$ and
$R(\mu)(3,3)=1$. Then $G_\nu$ is a subgroup of $SO(3)$ and we 
have proven \cite{HBEV2} that $(j,A)$ has spin tunes iff it has a $G_\nu$-normal form for some $\nu$.
\subsection{\label{ssec:DT} The Decomposition Theorem (DT)}
Choose $(E,l)$ and $(\hat{E},\hat{l})$, where  $E$ and $\hat{E}$ 
are Hausdorff and consider, in terms of 
functions $\beta\in\cC(E_\eta,\hat{E}_{\hat{\eta}})$,
the relation between fields which take values only in
$E_\eta:=l(SO(3);\eta)$ and $\hat{E}_{\hat{\eta}}:=\hat{l}(SO(3);\hat{\eta})$.
Then let $f\in\cC(\Td,E)$ take values only in $E_\eta$ whence, according
to (\ref{eq:FieldMap}), $f$ is mapped to $f'$ which takes 
values only in $E_\eta$ too. Of course $g\in\cC(\Td,\hat{E})$, defined by
$g(z):=\beta(f(z))$, takes values only in $\hat{E}_{\hat{\eta}}$ whence, according
to (\ref{eq:FieldMap}), $g$ is mapped to $g'$ which takes 
values only in $\hat{E}_{\hat{\eta}}$ too.
To relate $(E,l)$ and $(\hat{E},\hat{l})$
dynamically, we impose the condition on $\beta$ 
that $g'(z)=\beta(f'(z))$ for all $f$ and all $(j,A)$.
It is easy to show that this is equivalent to the condition
\be
 \hat{l}(r;\beta(x)) = \beta(l(r;x)) \; , \label{eq:betaSOThrMap}
\ee
i.e., that $\beta$ is an $SO(3)$-map.
We can now state the DT \cite{HBEV2}.
Firstly, assume an $SO(3)$-map $\beta$ exists, then 
$g'(z)=\beta(f'(z))$ for all $(j,A)$. Then if 
$f$ is an invariant $(E,l)$-field of $(j,A)$, $g$
is an invariant $(\hat{E},\hat{l})$-field 
of $(j,A)$. Secondly if, in addition, 
$\beta$ is a homeomorphism
then $f$ is an invariant $(E,l)$-field of $(j,A)$ iff $g$
is an invariant $(\hat{E},\hat{l})$-field of $(j,A)$; this allows us
to classify invariant fields up to homeomorphisms.

We now address the computation of the $\beta$'s.
One can show \cite{HBEV2} 
that an $SO(3)$-map $\beta$ exists  iff
 $H_\eta$ (see NFT above) 
 is conjugate to a subgroup
 of the isotropy group $\hat{H}_{\hat{\eta}}:=\{r\in SO(3) : \hat{l}(r;\hat{\eta})=\hat{\eta}\}$,
 i.e.\ an $r_0\in SO(3)$ exists so that $r_0 H_\eta r_0\tran \subset \hat{H}_{\hat{\eta}}$.
The proof is constructive by showing that $\beta$ can be defined by 
$\beta(l(r;x)) = \hat{l}(r\,r_0\tran ; \hat{x} )$.
Note that every $SO(3)$-map $\beta$ is a homeomorphism if 
$H_\eta$ and $\hat{H}_{\hat{\eta}}$ are conjugate, i.e.,
an $r_0\in SO(3)$ exists so that $r_0 H_\eta r_0\tran=\hat{H}_{\hat{\eta}}$.
In contrast, if $H_\eta$ and $\hat{H}_{\hat{\eta}}$ are not conjugate then
no $SO(3)$-map $\beta$ is a homeomorphism \cite{HBEV2}.

In summary, the DT enables one to classify and relate invariant fields
for the various choices of $\eta$ and $\hat{\eta}$
and it does so in terms of the functions $\beta$ and the
subgroups $H_\eta$ and $\hat{H}_{\hat{\eta}}$ of $SO(3)$.
The terminology DT refers to $E$ being ``decomposed'' into the
$E_\eta$ (similarly for $\hat{E}$).
%
%
\subsection{\label{ssec:IRT} The Invariant Reduction Theorem (IRT) and the Cross Section Theorem (CST)}
Throughout this section we consider a fixed $SO(3)$-space
$(E,l)$ and $x\in E$ and 
$f\in\cC(\Td,E)$ where $f(\Td)\subset E_x$.
Let $\check{\Sigma}_x[f]:=\bigcup_{z\in\Td}\;\lbrace z\rbrace\times 
{\cal R}_x(f(z))$
where, for $y\in E_x$, we
define ${\cal R}_x(y):=\lbrace r\in SO(3):l(r;x)=y \rbrace$. 
Moreover
$\check{\cal P}[j,A] \in \Homeo(\Td\times SO(3))$ is defined by
$\check{\cal P}[j,A](z,r):=(j(z),A(z)r)$. Note that 
$\check{\cal P}[j,A]:={\cal P}[SO(3),\ld,j,A]$ where
the $SO(3)$-space $(SO(3),\ld)$ is defined by $\ld(r';r):=r'r$.
Then the IRT \cite{He,HBEV1,HBEV2} states that $f$ is 
an invariant field iff $\check{\Sigma}_x[f]$ is 
$\check{\cal P}[j,A]$-invariant, i.e.,
$\check{\cal P}[j,A](\check{\Sigma}_x[f])=\check{\Sigma}_x[f]$.

The definitions of $\check{\Sigma}_x[f]$ and the name IRT
are suggested by bundle theory, as discussed 
in a separate section.
To sketch one-half of the IRT proof, note that if
$\check{\cal P}[j,A](z,r)=( j(z),A(z)r)\in\check{\Sigma}_x[f]$ then
$l(A(z)r;x)=f(j(z))$ whence, if $(z,r)\in\check{\Sigma}_x[f]$,
$l(A(z);f(z))=f(j(z))$ so that $f$ is invariant.

The IRT gives new insights into invariant fields.
Let $(z_0,y)\in \Td\times E_x$ and define
\be
\hat{\Sigma}_x[j,A,z_0,y] := 
 \bigcup_{n\in\Z} \check{\cal P}[j,A]^n(\lbrace z_0\rbrace\times {\cal R}_x(y)) \; .
\nonumber
\ee
Then a
corollary to the IRT states: 
If 
$\check{\Sigma}_x[f] = \Cl(\hat{\Sigma}_x[j,A,z_0,y])$,
where $\Cl$ indicates closure,
then $j$ is $TT(z_0)$ and
$f$ is an invariant field \cite{HBEV2}. 
Furthermore if $(z,r)\in \Cl(\hat{\Sigma}_x[j,A,z_0,y])$, 
then $f(z)=l(r;x)$ so that
$y$ explicitly determines the invariant field $f$ via 
the set $\Cl(\hat{\Sigma}_x[j,A,z_0,y])$.

Let $p_x[f]:\check{\Sigma}_x[f]\rightarrow\Td$ be defined by
$p_x[f](z,r):=z$. Hence $p_x[f]$ is continuous w.r.t. the 
subspace topology on $\check{\Sigma}_x[f]$.
One calls $\sigma\in\cC(\Td,\check{\Sigma}_x[f])$ 
a ``cross section'' of $p_x[f]$ if $p_x[f](\sigma(z))=z$. Then the CST states
that $p_x[f]$ has a cross section iff $T\in\cC(\Td,SO(3))$ exists such that
$f(z)=l(T(z);x)$ (the proof uses the fact 
that $\sigma(z):=(z,T(z))$
is a cross section). Note that, under the assumptions of the NFT, 
$p_x[f]$ has a cross section. The CST 
will be illustrated in Example 1
and will be tied with bundle theory
below.
%
%
%

We now state a partial converse to the above corollary.
Let $j$ be $TT(z_0)$, let $f$ be invariant and let
$p_x[f]$ have a cross section. Then 
\begin{eqnarray}
&&  \hspace{-5mm}
\check{\Sigma}_x[f] =
\Cl(\hat{\Sigma}_x[j,A,z_0,f(z_0)]) \; .
\label{eq:10.132nbaan}
\end{eqnarray}
Thus, as mentioned after the corollary, 
$f(z_0)$ explicitly determines the invariant field $f$ via 
the set $\Cl(\hat{\Sigma}_x[j,A,z_0,f(z_0)])$.
The fact that $f$ can be determined by the single value 
$f(z_0)$ is not surprising since
$j$ is $TT(z_0)$ 
and since the iteration
$f(j^{n+1}(z_0))=l(A(j^n(z_0))r;f(j^n(z_0)))$ 
gives $f$ on a dense subset of $\Td$ and, by continuity, everywhere.
In fact using the ``map'' version of Weyl's 
equidistribution theorem \cite{CFS}
one obtains an explicit
form of $f$ from $f(z_0)$ when $(E,l)=(\R^3,\lv)$.
In contrast, (\ref{eq:10.132nbaan}) is an alternative method of
obtaining an explicit
form of $f$ from $f(z_0)$ and it does so for arbitrary $(E,l)$. 
If $j$ is not $TT(z_0)$ then $f(z_0)$ does not 
necessarily determine $f$. For example let $j(z)=z$ and $A(z):=I_{3\times 3}$ then
every $f\in\cC(\Td, E)$ is an invariant $(E,l)$-field and hence
is not determined by $f(z_0)$. In contrast if $(E,l)=(\R^3,\lv)$ and
if $(j,A)$ are chosen such that only two ISF's $f,-f$ exist, then both
ISF's are uniquely determined by their value at $z_0$.
Note finally that for real storage rings there is also a well-known numerical
method of obtaining $f(z_0)$ and
constructing $f$ in terms of $f(z_0)$ when 
$(E,l)=(\R^3,\lv)$ and $(E,l)=(E_t,\lt)$
\cite{mont98,Ho,mv2000,Ba}.
%
%
%
%
%
\section{\label{sec:app} Application to the Examples}
\subsection{\label{ssec:vSpinApp}  
Example 1: $(E,l)=(\R^3,\lv)$}
We begin with the NFT. So let $x=(0,0,1)$ and
$f\in\cC(\Td,\R^3)$ be of the form $f(z)= \lv(T(z);x)=T(z)(0,0,1)$ where 
$T\in\cC(\Td,SO(3))$ whence $f$ is the third column of $T$.
Then the isotropy group is 
$H_x=\lbrace 
R(2\pi\nu)
:\nu\in[0,1)\rbrace
 =:SO(2)$
and the NFT states that
$T\tran(j(z))A(z)T(z) \in SO(2)$ for all $z\in\Td$ iff the third
column of $T$ is an ISF of $(j,A)$.
Note that as in \cite{BEH,HBEV2}
$T$ is called an ``invariant frame field'' (IFF) 
iff its third column is an ISF.
Thus our introduction of normal forms gives a new view on IFF's 
and generalizes it from the group $SO(2)$ to
an arbitrary subgroup of $SO(3)$.
We emphasize that the notions of IFF and 
ISF are tied to the group $SO(2)$.

To apply the DT we recall that the $E_x$ are 
the $\lv(SO(3);\lambda(0,0,1))$ which are spheres
of radius $\lambda$ centered at $(0,0,0)$ where $\lambda\in[0,\infty)$.
Thus, for topologically transitive $j$,
IPF's can be classified by applying the DT for the
case where $(E,l)=(\hat{E},\hat{l})=(\R^3,\lv)$ and
$\eta=\lambda(0,0,1),\hat{\eta}=\hat{\lambda}(0,0,1)$
with $\lambda,\hat{\lambda}\in[0,\infty)$.
One can show \cite{HBEV2} that
an $SO(3)$-map $\beta$ which is a homeomorphism only exists
if either $\lambda,\hat{\lambda}>0$ or $\lambda=\hat{\lambda}=0$.
Thus, for topologically transitive $j$,
there are only two classes of IPF's namely the class containing
the ISF's and the class consisting only of the zero field.

If $x=(0,0,1)$ and $f\in \cC(\Td,\R^3)$ with $|f|=1$
one observes, by the CST, that $p_x[f]$ 
has a cross section iff a $T\in\cC(\Td,SO(3))$ 
exists whose third column is $f$. Thus the CST gives another view on 
IFFs.
Using simple arguments from Homotopy Theory
one can also show \cite{He} that, if $d=1$, 
$p_x[f]$ has a cross section while for $d\geq 2$, 
$p_x[f]$ does not always have a cross section.
Thus, in light of (\ref{eq:10.132nbaan}), the case $d=1$ 
of the ISF conjecture
is exceptional.
\subsection{\label{ssec:tSpinApp} 
Example 2: $(E,l)=(\Et,\lt)$}
For brevity we only address the DT.
We call the fields $f\in\cC(\Td,\Et)$
polarization-tensor fields and so invariant $(\Et,\lt)$-fields
are invariant polarization-tensor fields (IPTF's).
Since $E_t$ is Hausdorff one can show \cite{HBEV2}, 
for topologically transitive $j$, that IPTF's 
can be classified by applying the DT for the
case where $(E,l)=(\hat{E},\hat{l})=(\Et,\lt)$ and
where $\eta$ and $\hat{\eta}$ are the diagonal matrices
$\eta=\diag(y_1,y_2,-y_1-y_2)$ and $\hat{\eta}
=\diag(\hat{y}_1,\hat{y}_2,-\hat{y}_1-\hat{y}_2)$ with
$y_1,y_2,\hat{y}_1,\hat{y}_2\in\R$.
One can also show \cite{HBEV2} that
an $SO(3)$-map $\beta$ which is a homeomorphism only exists
if $\eta$ and $\hat{\eta}$ have the same number of eigenvalues.
Thus, for topologically transitive $j$,
there are only three classes of IPTF's: the
class containing the invariant fields which have values in
$E_\eta$ where either $\eta=\diag(0,1,-1),\diag(1,1,-2)$, 
or $\diag(0,0,0)$.
Note that computing the underlying isotropy groups is easily done
by Linear Algebra since $l_t(r;M)$ is linear in $M$ (same for $\lv$).

We now apply the DT in the
case where  $(E,l)=(\R^3,\lv)$ and $(\hat{E},\hat{l})=(\Et,\lt)$ and
where $\eta=(0,0,1)$ and $\hat{\eta}=\diag(y,y,-2y)$
where $y\in\R$. We know from Example 1
that $H_\eta=SO(2)$ and one easily computes 
$\hat{H}_{\hat{\eta}}$ and finds that $H_\eta\subset\hat{H}_{\hat{\eta}}$ 
whence $\beta\in\cC(E_\eta,\hat{E}_{\hat{\eta}})$ defined by
$\beta(\lv(r;\eta)):=\lt(r; \hat{\eta} )$ is an
$SO(3)$-map. Easy computation shows that
$\beta(S)= y I_{3\times 3} -3ySS\tran$ where $S\in\R^3$ with $|S|=1$
so that if $f$ is an ISF of $(j,A)$ then, by the DT, the function
$g\in\cC(\Td,\Et)$, defined by
%
\be
g(z) :=\beta(f(z)) =  y I_{3\times 3} -3yf(z)f\tran(z) \,, \label{eq:ISFToITF}
\ee
is an IPTF of $(j,A)$.
This confirms the observation \cite{BV2} obtained by a different method.
\subsection{\label{ssec:vDeMatApp} 
Example 3: $(E,l)=(\EDensF,\lDensF)$}
We call every $\rho \in\cC(\Td,\EDensF)$
a spin-$1/2$ density matrix and since
$\alpha$ is a homeomorphism we can 
 write $\rho = \alpha \circ P$ which defines the polarization
 field $P\in\cC(\Td,\R^3)$.
With  (\ref{eq:StationaryEquation}) it follows that 
$\rho$ is an invariant field iff
$P$ is an IPF.
Thus the theorems about IPF's
(e.g., from the four main theorems) can be applied to invariant
$\rho$'s.

The spin-$1/2$ density matrix is the key to
 the statistical description of a bunch of spin-$1/2$ particles.
Assume for simplicity that the particle bunch is in equilibrium, so
that a stationary bunch density $\rho_{\text{eq}}(J)$ exists
where $J$ is the action-variable. Define
 $\rho_{\text{tot}}(n;z,J)$ $:=$
 $\rho_{\text{eq}}(J)\rho_{\text{spin}}(n;z,J)$ $=$ 
 $\rho_{\text{eq}}(J)\alpha(P(n;z,J))$ where each of the 
 $\rho_{\text{spin}}(n;\cdot,J)$ evolves under a $(j_J,A_J)$
as a spin-$1/2$ density matrix. 
Then $P(n;\cdot,J)$ moves as a polarization field
on the torus from turn $n$ to
 $n+1$. Note that $|P|\leq 1$.
Every ``physical observable'' 
${\cal O}\in\cC(\Td\times\Lambda,\C^{2\times 2})$ can be written as
${\cal O}(z,J) = h_0(z,J)+\sum_{i=1}^3 h_i(z,J)\sigma_i$, and its 
expectation value
  $\langle {\cal O} \rangle (n) $ at turn $n$ is defined by
\bea
 \langle {\cal O} \rangle (n) & := & 
    \int_{\Td\times\Lambda}dz\,dJ\, \trace \left(\rho_{\text{tot}} 
{\cal O} \right) \; ,
\label{eq:ExpectRho}
\eea
where $\Lambda\subset\R^d$ is the domain of $J$.
For example, for the spin observable 
${\cal O}(z,J):=\sigma_i$, the expectation value of ${\cal O}$
is the $i$-th component of the polarization vector of the
bunch at time $n$.
The choice $(\EDensF,\lDensF)$ and the above theory of
$\rho_{\text{tot}}$ follows from the semiclassical treatment
of Dirac's equation in terms of Wigner functions where the
particle-variables $z$ and $J$ are purely classical (see \cite{mont98}
and the references therein).
%
%
\subsection{\label{ssec:vXtDeMatApp} 
Example 4: $(E,l)=(\EDensB,\lDensB)$}
We call the fields $\rho \in\cC(\Td,\EDensB)$
spin-$1$ density matrices and since
$\alpha$ is a homeomorphism we can write 
$\rho(z)= \alpha(P(z),m(z))$ where the polarization field $P$
and the polarization-tensor field $m$ are
uniquely determined by $\rho$.
With  (\ref{eq:StationaryEquation}) one can show 
that $\rho$ is an invariant field iff
$P$ is an IPF and $m$ an IPTF of $(j,A)$.
Thus the theorems about IPF's and IPTF's can be applied to invariant
$\rho$'s.

We note that given $\rho$ one may construct in complete analogy to the 
spin-$1/2$ case a $\rho_{\text{tot}}$. The observables are
now continuous hermitian $3\times3$ matrix functions and one obtains the
analogy to (\ref{eq:ExpectRho}) by building 
$\rho_{\text{tot}}$ out of $\rho_{\text{eq}}$ and polarization fields 
and polarization-tensor fields.
In particular 
one can prepare an equilibrium bunch where the IPF's $P(0,\cdot,J)$
of $(j_J,A_J)$ are built up from the ISF's $p(0,\cdot,J)$
and where the IPTF's $m(0,\cdot,J)$
are built up from the $p(0,\cdot,J)$ via
(\ref{eq:ISFToITF}), and then $\rho_{\text{tot}}$
is completely determined by $\rho_{\text{eq}}$ and the 
$p(0,\cdot,J)$.
The choice $(\EDensB,\lDensB)$ and the theory of
$\rho_{\text{tot}}$ follows, as in the spin-$1/2$ case,
from the semiclassical treatment of Wigner functions.
\section{\label{sec:bundle} Underlying Bundle Theory}
While bundle-theoretic aspects play no role
in the above outline of our results 
we put it into that context here by following \cite{Fe,HK,Hu}.
See also \cite{He,HBEV2}. 

The ``unreduced'' principal bundle underlying our formalism
is a product principal $SO(3)$-bundle with base space $\Td$, i.e.,
it can be written as the $4$-tuple $(\Ed,p_d,\Td,L_d)$ where
$\Ed:=\Td\times SO(3)$
is the bundle space, $p_d\in\cC(\Ed,\Td)$ the
bundle projection, i.e., $p_d(z,r):=z$, and 
$(\Ed,L_d)$ the underlying $SO(3)$-space with 
$L_d:SO(3)\times \Ed\rightarrow \Ed$ defined by
$L_d(r;z,r'):=(z,r'r^t)$.
For every $(j,A)$, bundle theory gives
us a natural particle-spin map on $\Ed$ which turns out to be 
$\check{\cal P}[j,A]$. 
The reductions are those principal $H$-bundles which are
subbundles of the unreduced bundle such that their bundle space
is a closed subset of $\Ed$ and such that $H$ is closed. 
By the well-known Reduction Theorem \cite[Chapter 6]{Fe},
\cite[Chapter 6]{Hu}, every
$(\check{\Sigma}_x[f],p_x[f],\Td,L)$, for which $E$ is Hausdorff,
is a reduction where
$L$ is the restriction of $L_d$ to $H_x\times\check{\Sigma}_x[f]$
and conversely, every reduction is of this form.
By bundle theory, the natural particle-spin map on 
$\check{\Sigma}_x[f]$ for a given $(j,A)$
is that bijection on $\check{\Sigma}_x[f]$ which is
a restriction of $\check{\cal P}[j,A]$.
Clearly this function is a bijection iff
$\check{\Sigma}_x[f]$ is 
$\check{\cal P}[j,A]$-invariant and then the reduction is called
``invariant under $(j,A)$''.
Thus indeed the IRT deals with invariant reductions and it
states that a reduction is invariant under $(j,A)$ iff
$f$ is an invariant field.

The bundle-theoretic aspect of the CST follows from the simple fact that
the cross sections of $p_x[f]$ are the bundle-theoretic cross sections
of the reduction. Thus, by bundle theory,
$p_x[f]$ has a cross section iff the principal 
bundle $(\check{\Sigma}_x[f],p_x[f],\Td,L)$ is trivial, i.e.,
is isomorphic to a product principal bundle (this isomorphism
is used in proving (\ref{eq:10.132nbaan})).

Every $(E,l)$ in the formalism uniquely determines 
an ``associated bundle'' (relative to 
the unreduced bundle) which, up to bundle isomorphism, is of the
form $(\Td\times E,p,\Td)$ where $p(z,x):=z$.
Moreover, since $\check{\cal P}[j,A]$ is
the natural particle-spin map on $\Ed$, the association with
$(E,l)$ leads to a natural map on $\Td\times E$ which turns out
to be the particle-spin map ${\cal P}[E,l,j,A]$ of 
(\ref{eq:ParticleSpinMap}). In comparison, the matter 
fields in gauge field theories enter via associated bundles.

As a side aspect, the above mentioned reductions reveal a relation to
Yang-Mills Theory via the principal connections.
For example, in the presence of an IFF (see Example 1) we 
have an invariant $SO(2)$ 
reduction which has a cross section and describes planar spin motion.
Since this reduction is a smooth principal bundle, it has a well-defined class
 of principal connections leading via path lifting 
to parallel transport motions which,
remarkably, reproduce
 the form of the well-known T-BMT equation, and thus in discrete time give 
us ${\cal P}[\R^3,\lv,j,A]$.
These aspects will be extended to nonplanar spin motion \cite{HBEV3}. 
\section*{Acknowledgments}
 Work supported by DOE under DE-FG02-99ER41104
 and by DESY.

%

\end{document}